\documentclass[prl,twocolumn,showpacs,preprintnumbers,aps,nofootinbib]{revtex4-1}
\usepackage{graphicx}
\usepackage{dcolumn}
\usepackage{bm}
\usepackage{amsmath,amssymb,amsfonts}
\usepackage{latexsym}
\usepackage{color}

\begin{document}


\title{\boldmath
Constraining a Lighter Exotic Scalar via Same-sign Top
}

\author{Wei-Shu Hou, Masaya Kohda and Tanmoy Modak}
\affiliation{Department of Physics, National Taiwan University, Taipei 10617, Taiwan}
\bigskip

\date{\today}

\begin{abstract}
It was shown recently that, in two Higgs doublet models 
without $Z_2$ symmetry, extra Yukawa couplings such as 
$\rho_{tc}$, $\rho_{tt}$ can fuel enough $CP$ violation for
electroweak baryogenesis (EWBG).
We revisit an old proposal where a pseudoscalar $A^0$ has mass
between $t\bar c$ and $t\bar t$ thresholds.
With $\rho_{tt}$ small, it evades $gg \to A^0 \to h^0(125)Z$ constraints,
where approximate alignment also helps.
We find this scenario with relatively light $A^0$ is not yet ruled out,
and $cg \to tA^0 \to tt\bar c$ can probe sizable $\rho_{tc}$ at the LHC,
giving access  to the second mechanism of EWBG provided by such models.
\end{abstract}

\maketitle

\paragraph{Introduction.---}

The absence of clear signs of New Physics (NP) at the Large Hadron Collider (LHC)
motivates one to think in less conventional ways.
In searching for extra Higgs bosons, 
it is common to assume a two-Higgs doublet model (2HDM) 
with a softly-broken $Z_2$ symmetry that implements 
the Natural Flavor Conservation (NFC) condition~\cite{Glashow:1976nt}.
Such $Z_2$ symmetries ensure that each type of charged fermions couples
to a single Higgs doublet, which thereby excludes 
the possible existence of extra Yukawa matrices.
However, given that we still do not understand the origin of the Yukawa sector,
the $Z_2$ assumption may appear too strong.
With a first doublet established since 2012~\cite{PDG:2018}, 
it seems imperative that we should use experimental {\it data} 
to constrain possible extra Yukawa couplings.

The 2HDM  without the $Z_2$ symmetries 
offers extra Yukawa couplings
that induce flavor-changing neutral Higgs (FCNH) interactions at tree level.
In particular, the Yukawa matrix element $\rho_{tc}$, 
the $tc S^0$ ($S^0 = H^0, A^0$) coupling, 
may be large because it involves the heaviest quark, top.
The $S^0 \to t \bar c, \bar tc$ width may well exceed $S^0 \to b\bar b$,
and could be the dominant decay mode for $m_{S^0}$ lying
between the $t\bar c$ and $t\bar t$ thresholds.
Through the $cg \to t S^0 \to tt\bar{c}$ process, one may have 
same-sign top-quark pair production at the LHC.

Such processes were first studied for the pseudoscalar boson $A^0$
about two decades ago~\cite{Hou:1997pm}, in the scenario 
that $m_{A^0} < m_{H^+} +M_W$, $m_{h^0/H^0}+M_{Z}$.
Assuming the Higgs sector is $CP$-conserving,
two-body $A^0$ decays are then limited to fermionic final states at tree level.
Invoking the Cheng-Sher ansatz~\cite{Cheng:1987rs} to allow for 
sizable $tcA^0$ coupling, $A^0 \to t\bar c$ would be the dominant decay 
for $m_{A^0}$ between the $t\bar c$ and $t\bar t$ thresholds.
For $200~{\rm GeV} \lesssim m_{A^0} \lesssim 2m_t$, 
it was advocated that $cg \to tA^0 \to tt\bar{c}$~\cite{ttc-recent} 
is a promising process to probe for the $tcA^0$ coupling.

\begin{figure*}[htbp!]
\center
\includegraphics[width=.32 \textwidth]{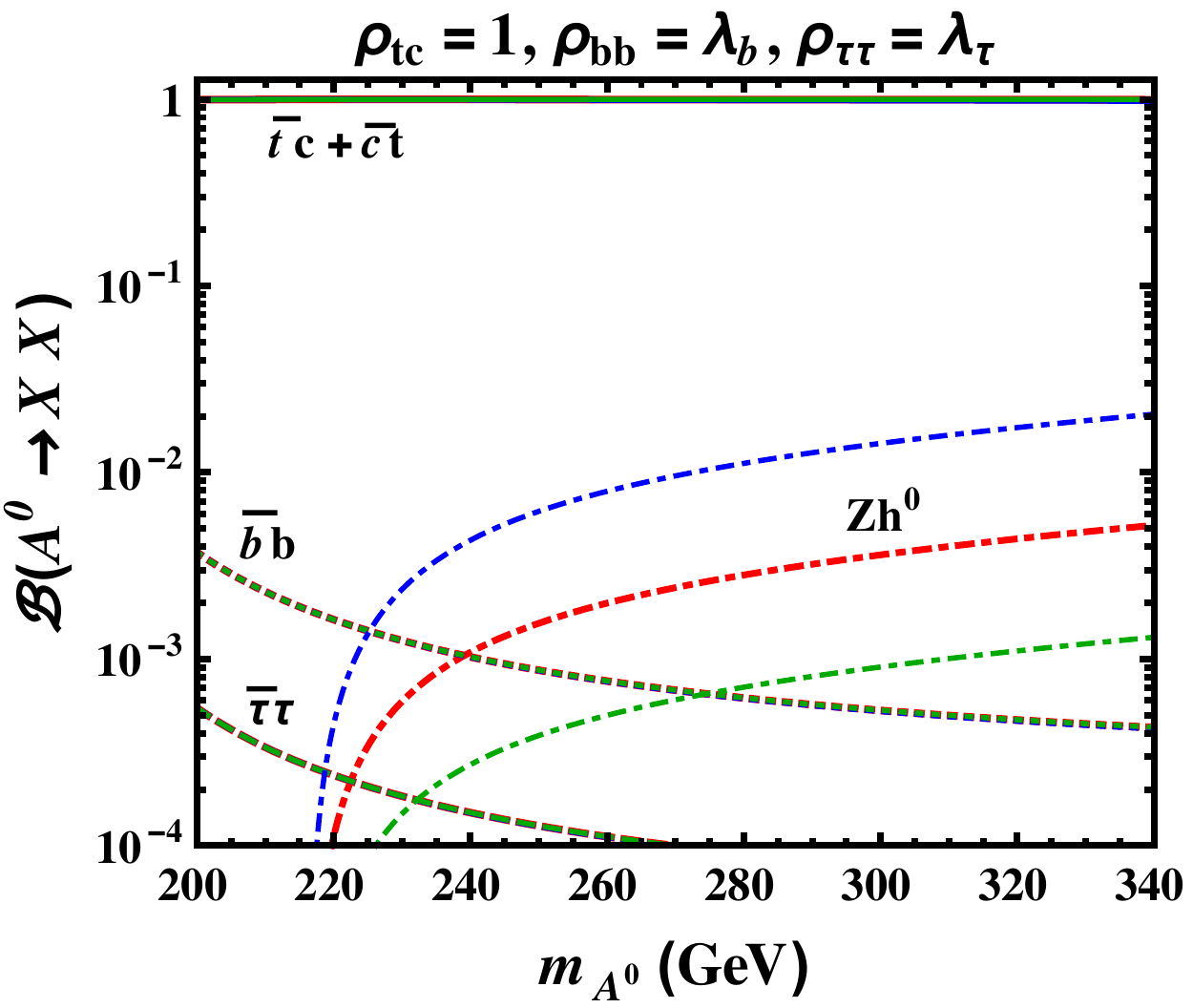}
\includegraphics[width=.32 \textwidth]{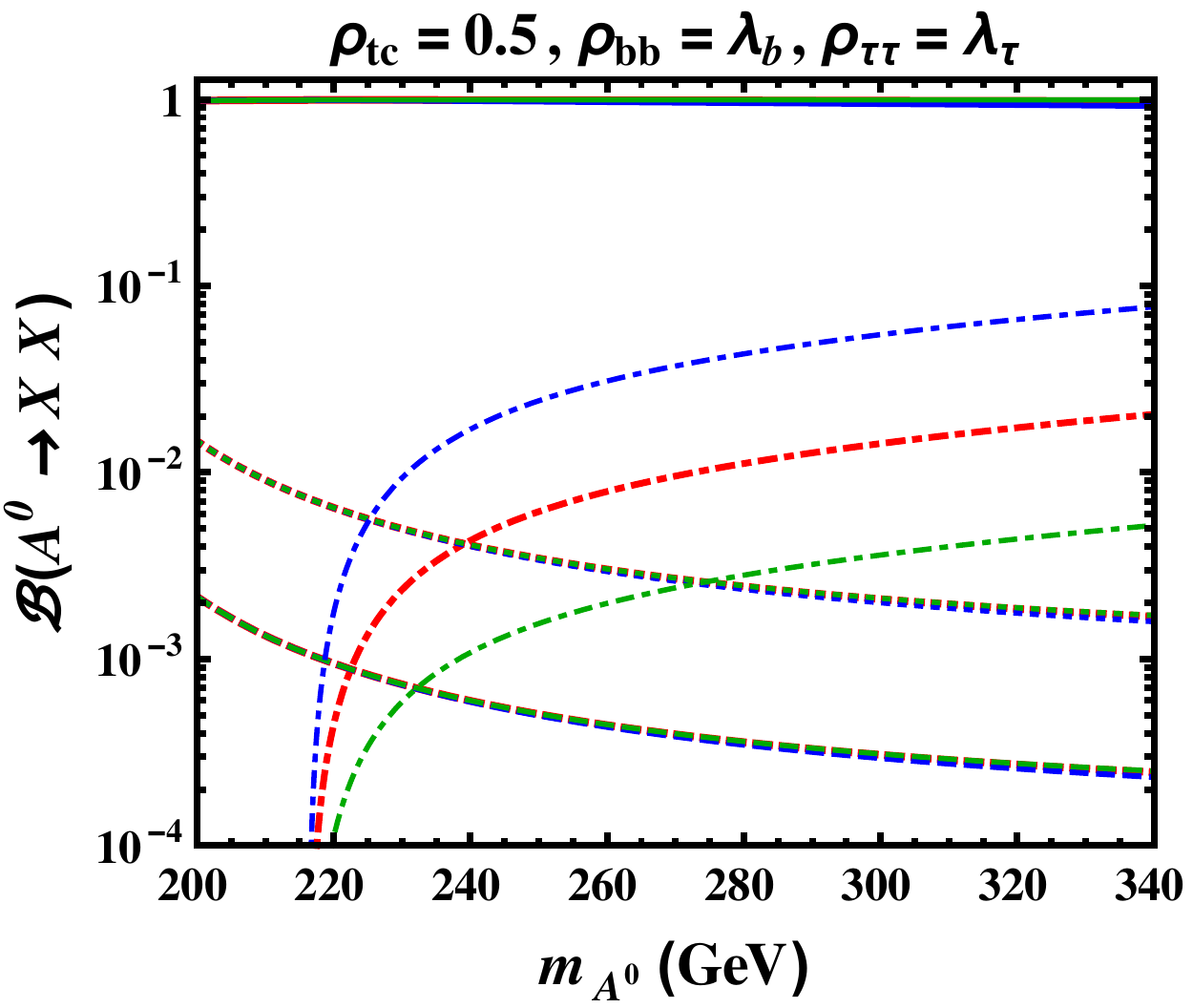}
\includegraphics[width=.32 \textwidth]{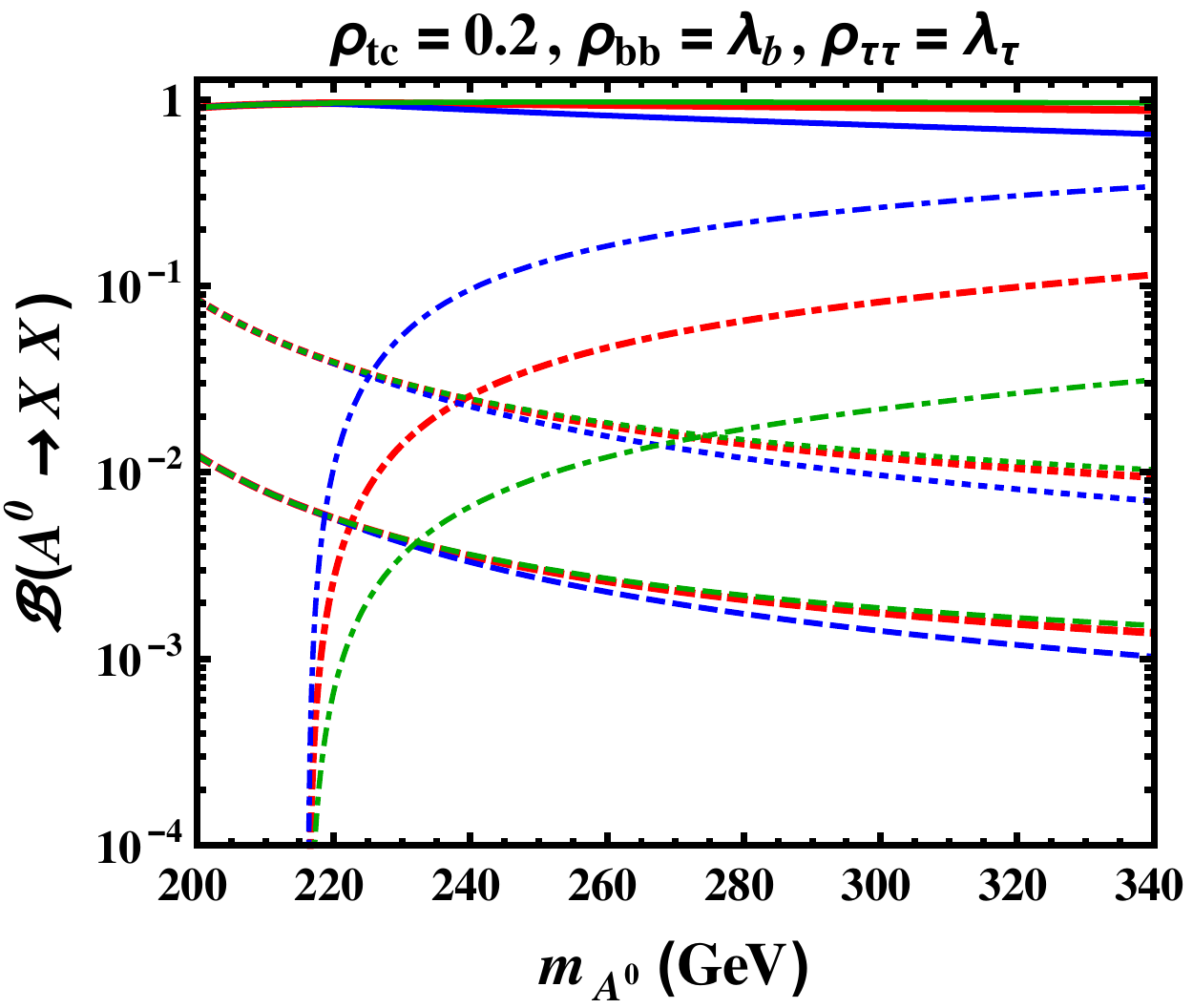}
\caption{
For $\rho_{tc} = 1$ [left], $0.5$ [middle] and $0.2$ [right],
branching ratios of $A^0$ decay to $t \bar c +\bar tc$ (solid), 
$b \bar b$ (dotted), $\tau^- \tau^+$ (dashed) and $h^0Z$ (dot-dashed)
with $c_\gamma =0.2$ (blue), $0.1$ (red) and $0.05$ (green).
}\label{brplot}
\end{figure*}

The case for sizable $\rho_{tc}$ was strengthened recently from cosmological concerns.
It was shown~\cite{Fuyuto:2017ewj} that in 2HDM without $Z_2$ symmetry,
top/charm transport can generate enough  $CP$ violation (CPV)
for successful electroweak baryogenesis (EWBG)~\cite{Kuzmin:1985mm}
during the electroweak phase transition (EWPT), which can be of 
strong first order if Higgs quartic couplings are $\mathcal{O}(1)$~\cite{Bochkarev:1990fx}.
The extra flavor-diagonal complex Yukawa coupling $\rho_{tt}$ is 
more efficient in generating the CPV, 
but an $\mathcal{O}(1)$ $\rho_{tc}$ with a large phase
can still realize EWBG even if $\rho_{tt}$ is rather small.

With the advent of LHC data and the renewed motivation, 
we revisit the study of Ref.~\cite{Hou:1997pm}.
With the discovered~\cite{h125_discovery} 125 GeV boson $h^0$~\cite{h125}
being quite consistent with the Higgs boson of the Standard Model
(SM)~\cite{Khachatryan:2016vau}, the scenario needs some update.
In particular, 
assuming $m_{A^0} < m_{h^0}+M_{Z}$ would preclude most of the mass range of interest,
hence 
must be dropped.
This opens up the $A^0 \to h^0Z$ decay, 
which is suppressed by proximity to the alignment limit~\cite{Gunion:2002zf},
as implied~\cite{cos_b-a} by LHC data.
To simplify our analysis, we focus on $A^0$ and assume it to be considerably lighter than 
$H^0$ and $H^+$, 
which is stronger than the original assumption of $m_{A^0} < m_{H^+} +M_W$, $m_{H^0}+M_{Z}$
of Ref.~\cite{Hou:1997pm}.
We discuss possible effects of $cg \to tH^0 \to tt\bar{c}$ near the end of this paper.

In this paper, we assume both top quarks decay semileptonically 
in $cg \to tA^0 \to tt\bar{c}$, leading to the signature of 
same-sign dilepton with jets and missing energy (SS2$\ell$).
We take an agnostic view on the FCNH coupling $\rho_{tc}$, 
treating it as a free parameter.
We survey existing LHC data pertaining to the SS2$\ell$ signature to constrain $\rho_{tc}$,
then study the discovery/exclusion potential at the LHC.
It is intriguing that the light $A^0$ scenario of Ref.~\cite{Hou:1997pm} 
is still allowed by existing data, and that future LHC data 
can probe $\rho_{tc}$ values relevant for EWBG.


\paragraph{ Framework and constraints.---}

We assume the Higgs potential is $CP$-conserving~\cite{Gunion:2002zf}.
Without a $Z_2$ symmetry, the coupling of $A^0$ to fermions is~\cite{Davidson:2005cw}
\begin{align}
\frac{i}{\sqrt{2}} \sum_{F = U, D, L}
 {\rm sgn}(Q_F) \rho^F_{ij} \bar F_{iL} F_{jR} A^0  +{\rm h.c.},
\end{align}
where $i,j =1,2,3$ are generation indices, ${\rm sgn}(Q_F) = +1$ ($-1$) 
for $F = U$ ($F=D, L$), and $\rho^F$ are general $3\times 3$ complex matrices.
For the $tc A^0$ couplings of interest,  
$\rho^U_{23}\equiv \rho_{ct}$ and $\rho^U_{32}\equiv \rho_{tc}$.
$B$ physics sets stringent limits on $\rho_{ct}$~\cite{Altunkaynak:2015twa},
while $\rho_{tc}$ is only mildly constrained~\cite{Crivellin:2013wna},
depending on $m_{H^+}$.
In our study, we set $\rho_{ct} = 0$ and vary $\rho_{tc}$ within $|\rho_{tc}| \lesssim 1$.

Other couplings can affect our study through the $A^0 \to t\bar c$ branching ratio: 
important ones are $\rho^D_{33} \equiv \rho_{bb}$ and $\rho^L_{33} \equiv \rho_{\tau\tau}$,
where each may be as large as the corresponding SM Yukawa coupling
$\lambda_f = \sqrt{2} m_f/v$ with $v \simeq 246$ GeV, leading to 
$A^0 \to b\bar b,\, \tau^+\tau^- $.
A nonzero $\rho^U_{33} \equiv \rho_{tt}$ induces $A^0 \to t\bar t$,
which is forbidden below the $t\bar t$ threshold, but 
can still generate $A^0 \to gg, \gamma\gamma$ via the triangle loop.
Finally, $A^0 \to h^0Z$ can occur via the gauge coupling
\begin{align}
 \frac{g_2 \cos\gamma}{2\cos\theta_W}Z_\mu(h^0\partial^\mu A^0 
 -A^0 \partial^\mu h^0),
\end{align}
where 
$\cos\gamma$~\cite{Hou:2017hiw} is the $CP$-even scalar mixing,
usually~\cite{Gunion:2002zf} denoted as $\cos(\alpha -\beta)$ 
in models with $Z_2$ symmetry.


We consider the mass range~\cite{Hou:1997pm}  of
$200~{\rm GeV} < m_{A^0} < 340~{\rm GeV}$ throughout this paper.
Taking $\rho_{bb} = \lambda_b$ and $\rho_{\tau\tau} = \lambda_\tau$ for illustration, 
we present $A^0$ decay branching ratios in Fig.~\ref{brplot}
for $\rho_{tc}=1$ (left), $0.5$ (middle) and $0.2$ (right),
with all other $\rho^F_{ij}$ set to zero.
In each panel, results for three different $c_\gamma \equiv \cos\gamma$ values are shown:
$0.2$ (blue), $0.1$ (red) and $0.05$ (green).
We see that $A^0 \to t\bar c, \bar tc$ are the dominant decay modes in all the cases
and $\mathcal{B}(A^0\to t \bar c + \bar tc) \sim 1$ for $\rho_{tc} \gtrsim 0.5$.
For $\rho_{tc} = 0.2$, other decay modes can become sizable, e.g. 
$\mathcal{B}(A^0\to h^0Z) \gtrsim 0.2$ for $c_\gamma =0.2$ and 
$270~{\rm GeV} \lesssim m_{A^0} < 340$ GeV; however, 
$\mathcal{B}(A^0\to t \bar c + \bar tc) > 60\%$ in all cases.

For nonzero $\rho_{tt}$, $gg \to A^0$ via the top loop 
makes $A^0 \to h^0Z$ search at the LHC relevant.
In the $A^0$ mass range of interest and for $|\rho_{tt}| \sim \lambda_t \sim 1$,
recent searches by ATLAS~\cite{Aaboud:2017cxo} and CMS~\cite{CMS:2018xvc},
both using $h^0 \to b\bar b$ with $\sim 36$ fb$^{-1}$ data at 13 TeV,
are sensitive to $\mathcal{B}(A^0\to h^0Z)$ at percent level.
Furthermore, diphoton resonance search can also become relevant
for  $|\rho_{tt}| \sim \lambda_t$.
For simplicity, we assume $|\rho_{tt}| \ll 1$ to suppress  $gg \to A^0$.
Note that one may still have the $\rho_{tc}$-driven EWBG~\cite{Fuyuto:2017ewj} 
even in this case. 
In the following analyses, we set all $\rho^F_{ij} = 0$  except for $\rho_{tc}$, 
and assume the alignment limit where $c_\gamma = 0$,
so that $\mathcal{B}(A^0\to t \bar c + \bar tc) = 1$ always holds.

\begin{figure*}[htbp!]
\center
\includegraphics[width=.4 \textwidth]{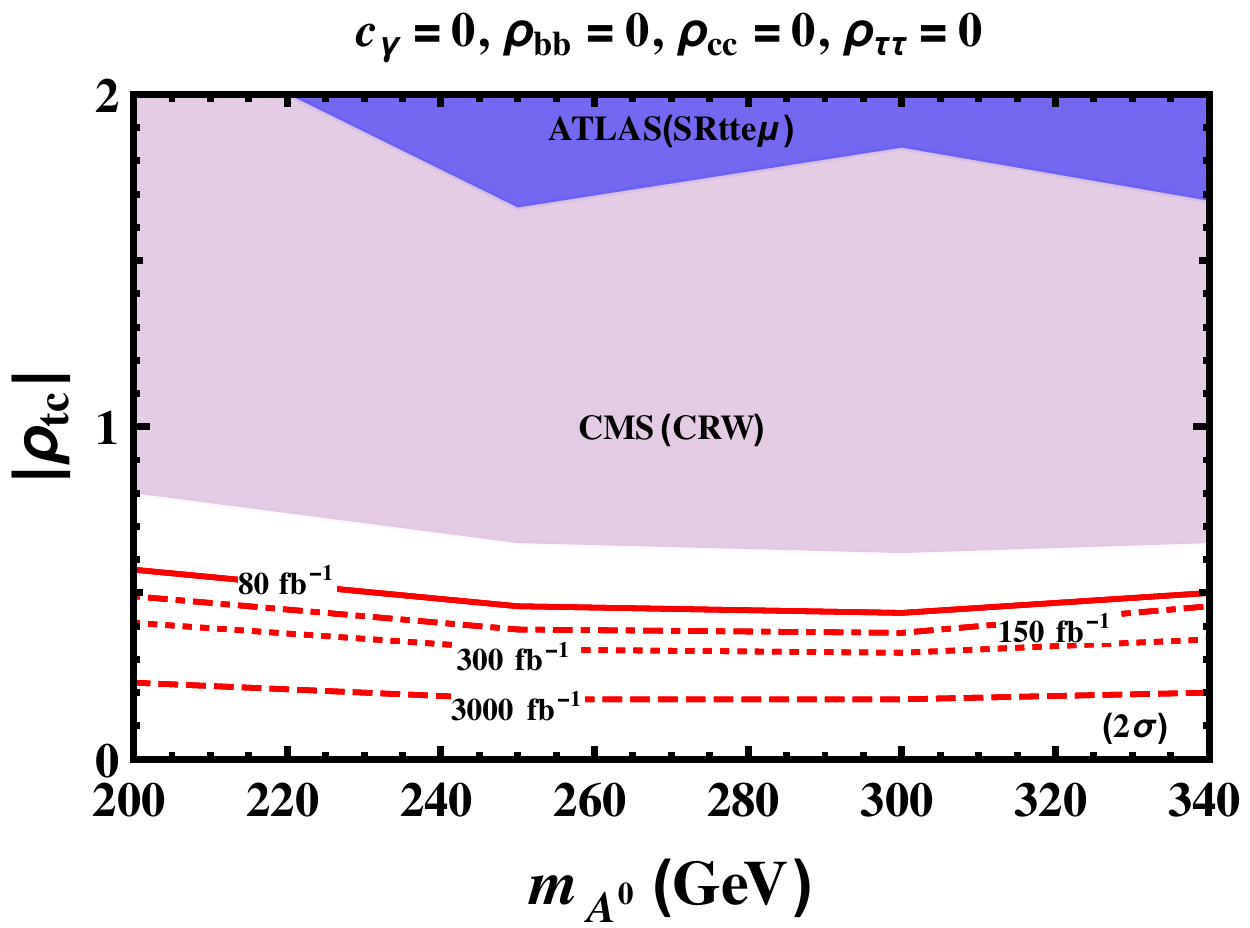}
\includegraphics[width=.4 \textwidth]{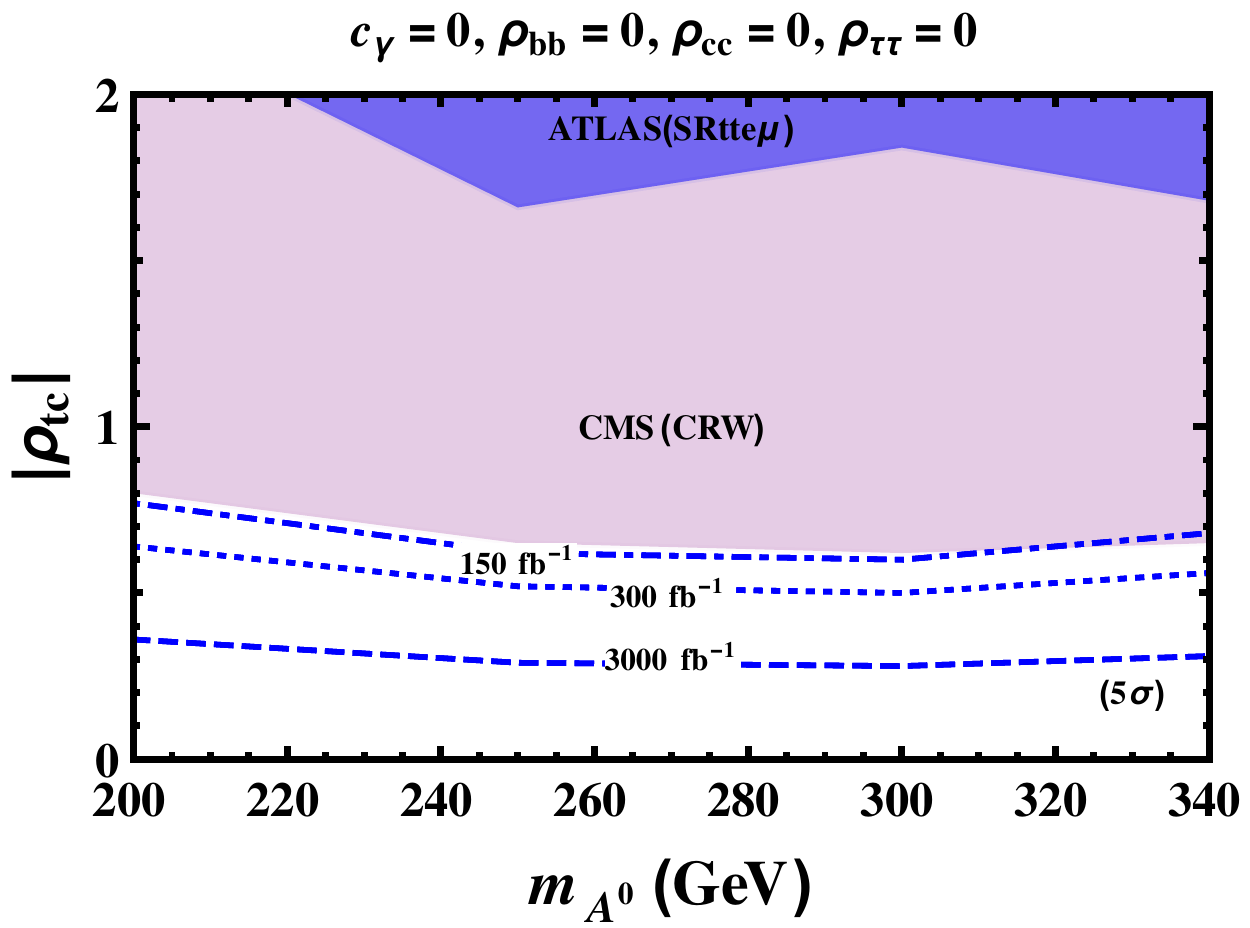}
\caption{
2$\sigma$ exclusion limits [left] and 5$\sigma$ discovery reaches [right] 
for $|\rho_{tc}|$ by SS2$\ell$ with various integrated luminosities at 14 TeV.
In both panels, $2\sigma$ excluded regions by existing data are overlaid:
SRtt$e\mu$ of $cc \to tt$ search by ATLAS (blue shaded) and 
CRW of $t\bar t t \bar t$ search by CMS (purple shaded).
See text for details.
}\label{exclusion}
\end{figure*}

A nonzero $\rho_{tc}$ induces $cg \to tA^0$, followed by 
$A^0 \to t\bar{c}, \bar t c $ in our setup. 
The $t\bar t c$ final state would be obscured by QCD production of $t\bar t$, 
but $tt\bar{c}$ with semileptonically decaying top gives clean 
same-sign dilepton signature, and should provide an excellent probe for $\rho_{tc}$.
Although there are no dedicated  searches for a new boson in such a process, 
one may utilize existing results for NP searches 
in same-sign dilepton final states to constrain $\rho_{tc}$.

Surveying literature, we find two relevant experimental results.
The first is the search by ATLAS~\cite{Aad:2015gdg} 
for $qq \to tt$ ($q = u$ or $c$) mediated by $t$-channel scalar 
$H$ exchange with $tqH$ coupling, using 20.3 fb$^{-1}$ data at 8 TeV.
The other is the search for SM production of four top quarks ($t\bar t t\bar t$)
by CMS~\cite{Sirunyan:2017roi}, using 35.9 fb$^{-1}$ data at 13 TeV.
We note that searches for supersymmetry in similar event topologies 
typically require missing energies that are too large for our purpose.
The requirement can be relaxed with $R$-parity violation,
for example a search by ATLAS~\cite{Aaboud:2017dmy} for 
squark pair production in $pp \to \tilde d_R \tilde d_R \to \bar t \bar t \bar b \bar b$ 
or $\bar t \bar t \bar s \bar s$. The selection cuts, however, 
 are still too strong to give meaningful constraints on $\rho_{tc}$.

The ATLAS $qq\to tt$ search~\cite{Aad:2015gdg}, depending on lepton flavor,  
defines three signal regions (SRs), where we find SRtt$e\mu$
($e\mu$ final state from both tops decaying semileptonically)
gives the best limit on $\rho_{tc}$. 
%
%
On the other hand, based on the number of leptons, jets or $b$-tagged jets, 
the CMS $t\bar t t\bar t$ search~\cite{Sirunyan:2017roi} 
defines eight SRs plus two control regions for background.
We find CRW, the control region for $t\bar tW$ background, 
gives the best limit.
Note that CMS has an earlier study~\cite{Sirunyan:2017uyt} of $t\bar t t\bar t$
based on the same dataset, 
but Ref.~\cite{Sirunyan:2017roi} has better 
optimization to enhance signal sensitivity.

Let us take a closer look.
The selection criteria for SRtt$e\mu$ 
of the ATLAS $qq\to tt$ search~\cite{Aad:2015gdg}
requires an event to contain a positively-charged $e\mu$ pair and
two to four jets with at least one $b$-tagged.
The transverse momenta, $p_T$, of both leptons should be $> 25$ GeV,
the pseudo-rapidity, $|\eta|$, of electrons (muons) should be $< 2.47~(2.5)$,
while all jets are required to have $p_T > 25$ GeV and $|\eta|< 2.5$. 
The azimuthal separation between the two leptons should be $\Delta\phi_{\ell\ell} > 2.5$. 
Finally, the missing $p_T$, or $p_T^{\rm miss}$, should be $>40$ GeV,
and the scalar sum, $H_T$, of all jet and lepton $p_T$s 
is required to be $> 450$ GeV.
With these selection cuts, ATLAS reports 5 observed events, 
with expected background at
{$7.5 \pm 1.3\, ({\rm stat}) \pm 2.5\, ({\rm syst})$}. 

CRW of the CMS $t\bar t t\bar t$ search~\cite{Sirunyan:2017roi} is defined 
to contain two same-sign leptons, two to five jets with two of them $b$-tagged.
The selection cuts are as follows.
Leading (subleading) lepton $p_T$ should be $> 25$ (20) GeV. 
The $|\eta|$ of electrons (muons) should be $< 2.5$ ($2.4$), while $< 2.4$ for all jets.
The events are selected if $p_T$ of ($b$-)jets satisfy 
any of the following three conditions~\cite{info-Jack}:
(i) both $b$-jets satisfy $p_T>40$ GeV;
(ii) one $b$-jet has $p_T> 20$ GeV and the other $b$-jet satisfies $20 < p_T< 40$ GeV,
with a third jet having $p_T > 40$ GeV;
(iii) both $b$-jets satisfy $20< p_T< 40$ GeV, with two extra jets 
each having $p_T > 40$ GeV. 
$H_T$, defined here as the scalar sum of $p_T$ of all jets~\cite{HT-def}, 
should be $> 300$ GeV, while $p_T^{\rm miss} > 50$ GeV.
In order to reduce the Drell-Yan background with a charge-misidentified electron,
events with same-sign electron pairs with $m_{ee}$ below 12 GeV are rejected.
With these selection cuts, CMS reports 86 observed events in CRW,
where the expected number of total events 
(SM backgrounds plus $t\bar t t\bar t$) is $85.6 \pm 8.6$.

The $\rho_{tc}$-driven process $pp\to t A^0 \to t t \bar c$
(with $cc \to tt$ via $t$-channel $A^0$ exchange and non-resonant $cg \to tt\bar c$ included), 
with both top quarks decaying semileptonically, 
contributes to SRtt$e\mu$ and CRW.
We estimate this contribution for $\rho_{tc} = 1$, then scale it 
by $|\rho_{tc}|^2$ assuming a narrow $A^0$ width and 
$\mathcal{B}(A^0\to t \bar c) = 0.5$, which is ensured by
$c_\gamma =0$ and $\rho^F_{ij}=0$ except $\rho_{tc}$. 
We demand the sum of the expected number of events for the $tA^0$ and 
SM contributions to agree with the number of the observed events within 
$2\sigma$ uncertainty for the SM expectation.
Simplifying by assuming Gaussian~\cite{excl-poisson} behavior for the latter,
the $2\sigma$ exclusion limits 
obtained for SRtt$e\mu$ and CRW are displayed in Fig.~\ref{exclusion} 
as the blue and purple shaded regions, respectively.

In calculating these limits, we generate the signal events by 
MadGraph5\_aMC@NLO~\cite{Alwall:2014hca} (denoted as MadGraph5\_aMC) at leading order (LO) 
with default parton distribution function (PDF) set NN23LO1~\cite{Ball:2013hta}, 
interfaced with PYTHIA~6.4~\cite{Sjostrand:2006za} for showering and hadronization, 
and MLM matching~\cite{Alwall:2007fs} prescription for matrix element and parton shower merging.
The event samples are fed into Delphes~3.4.0~\cite{deFavereau:2013fsa} for fast detector simulation, 
following CMS based detector analysis for CRW and ATLAS based for SRtt$e\mu$. 
The effective model is implemented in FeynRules~\cite{Alloul:2013bka}.
We utilize the default $p_T$ and $\eta$-dependent $b$-tagging efficiency implemented in Delphes card, while
take rejection factors $5$ and $137$ for $c$-jets and light-jets, respectively~\cite{ATLAS:2014ffa}.

Inspecting the exclusion limits in Fig.~\ref{exclusion},
the limits by SRtt$e\mu$~\cite{Aad:2015gdg} is above $|\rho_{tc}| \sim 1.5$ in general.
Since the selection cuts become a bit too strong for $m_{A^0}\lesssim 240 ~\mbox{GeV}$, 
the constraint on $|\rho_{tc}|$ becomes weaker.
On the other hand, CRW~\cite{Sirunyan:2017roi} rules out $|\rho_{tc}|\gtrsim 0.75$ for $m_{A^0} \lesssim 230$ GeV, 
and $|\rho_{tc}|\gtrsim 0.65$ for $250 ~\mbox{GeV}\lesssim m_{A^0} \lesssim 340~\mbox{GeV}$. 
Thus, the CRW data already probe the $\rho_{tc}$-driven EWBG.
%

%
%

%

\paragraph{LHC prospects for same-sign top.---}

Although the existing experimental results can be used to set meaningful constraints,
they are not optimized for $cg \to tA^0 \to tt\bar c$ search.
In this section, we explore the potential of LHC to exclude or discover
$pp\to t A^0 \to t t\bar c$ in the SS2$\ell$ signature,
which we define as an event containing same-sign dilepton, at least three jets 
with at least two $b$-tagged, and missing transverse momentum.
We follow our previous study~\cite{Kohda:2017fkn} for $m_{S^0} > 2m_t$,
but add to the signal the contribution from $cc \to tt$ via $t$-channel $A^0$ exchange.
SM background processes are $t\bar t Z$, $t\bar t W$,
$tZ +$ jets, $3t + j$, $3t + W$, $4t$, and $t\bar t h$. 
The $Z/\gamma^*+$ jets or $t\bar t+$ jets processes would also contribute 
if the charge of a lepton gets misidentified (charge- or $Q$-flip), with the probability
$2.2\times 10^{-4}$~\cite{ATLAS:2016kjm,Alvarez:2016nrz}.
Furthermore, the CMS study~\cite{Sirunyan:2017uyt} for similar final state 
with slightly different cuts finds ``nonprompt''
backgrounds to be significant ($\sim1.5$ times the $t\bar{t}W$ background).
These backgrounds are not properly modeled in Monte Carlo simulations,
we thus simply add to the overall background a nonprompt component 
that is 1.5 times the $t\bar t W$ after selection cuts.

The signal and background events are generated at LO via MadGraph5\_aMC for $\sqrt{s}=14$ TeV,
and we follow the same procedure as in previous section for showering, hadronization and matching,
and adopt ATLAS based detector card in Delphes.
We set $c_\gamma=0$ and $\rho^F_{ij} = 0$ except for $\rho_{tc}$.
For cut based analysis and details of the QCD corrections for the
background processes, we follow Ref.~\cite{Kohda:2017fkn}.
Note that we do not include QCD corrections for signal.
The signal (for different $m_{A^0}$) and background cross sections after
the selection cuts are summarized in Tables~\ref{signal} and~\ref{backg_ssll}, respectively.


\begin{table}[t]
\centering
\begin{tabular}{c |c c c c }
\hline\hline
                      $m_{A^0}$ (GeV)                 &  Cross section (fb)      \\
\hline 
                       200                 & 0.998                \\
                       250                 & 1.534               \\
                       300                 & 1.666                     \\
                       340                 & 1.300               \\

\hline
\hline
\end{tabular}
\caption{
Signal cross sections of SS2$\ell$ for $\rho_{tc}=1$ at 14 TeV LHC.
The event selection cuts are imposed.
}\label{signal}
\end{table}

\begin{table}[t]
\centering
\begin{tabular}{c |c c c c }
\hline\hline
                      Backgrounds                 &  Cross section (fb)      \\
\hline 
                       $t\bar{t}Z$                 & 0.04                \\
                       $t\bar{t}W$                 & 0.72               \\
                       $tZ+$jets                   & 0.001                     \\
                       $3t+j$                      & 0.0002               \\
                       $3t+W$                      & 0.0004               \\
                     {$t\bar t h$}                & {0.024}                      \\
                       $4t$                        & 0.04                  \\
                       $Q$-flip                    & 0.04                     \\
\hline
\hline
\end{tabular}
\caption{Same as Table.~\ref{signal} but for backgrounds.}
\label{backg_ssll}
\end{table}

In order to project the exclusion limit and the discovery potential for $n$ observed events, 
we utilize the significance expression~\cite{excl-poisson,Craig:2016ygr}
\begin{align}
Z(x|n)=\sqrt{-2\ln\frac{L(x|n)}{L(n|n)}},
\end{align}
and likelihood function given by Poisson counting experiment $L(x|n) =  e^{-x}x^n/n!$, 
where $x$ is either the number of events predicted by the background only hypothesis $b$,
 or signal plus background hypothesis $s+b$.
We demand $Z(s+b|b) \geq 2$ for $2\sigma$ exclusion,
while $Z(b|s+b) \geq 5$ for $5\sigma$ discovery.
Using the 
cross sections in Tables~\ref{signal} 
and~\ref{backg_ssll}, we overlay the exclusion (red) and discovery (blue) contours for SS2$\ell$ in the left and right 
panels of Fig.~\ref{exclusion}, respectively. For simplicity,   
we interpolate these contours from $m_{A^0}$ values given in Table~\ref{signal}.

We assume $\sqrt{s}=14$ TeV for exclusion and discovery contours 
in Fig.~\ref{exclusion}, and give three different integrated luminosities: 
150 fb$^{-1}$ (dot-dashed), 300 fb$^{-1}$ (dotted), 3000 fb$^{-1}$ (dashed).
We add $80$ fb$^{-1}$ (red solid) to the left panel to illustrate 
how well current data can exclude $|\rho_{tc}|$ with a dedicated analysis, 
while 150 fb$^{-1}$ is the target luminosity for Run-2,
although the numbers are for $14$, rather than 13 TeV. 
We find that, with 80 (150) fb$^{-1}$ data, 
$|\rho_{tc}|\gtrsim 0.6~(0.5)$ could be excluded  for 
$m_{A^0}\lesssim 230\,\mbox{GeV}$, 
and $|\rho_{tc}|\gtrsim 0.5~(0.4)$ 
for $240\,\mbox{GeV}\lesssim m_{A^0}\lesssim 340$ GeV,
while 3000 fb$^{-1}$ data can probe $|\rho_{tc}|$ down to 0.2.
One would certainly need larger $|\rho_{tc}|$ for discovery. 
For instance, discovery is possible for $|\rho_{tc}| \gtrsim 0.6~(0.4)$
with $300~(3000)$ fb$^{-1}$ for $m_{A^0} \lesssim 230~\mbox{GeV}$,
while reaching down to $|\rho_{tc}|\sim 0.5~(0.3)$ 
for $240~\mbox{GeV}\lesssim m_{A^0}\lesssim 340$ GeV.

\paragraph{Discussion and Conclusion.---}

In a previous work~\cite{Kohda:2017fkn}, we have studied
same-sign dilepton and triple top signatures that 
arise from $cg \to tA^0,\, tH^0 \to tt\bar c,\, tt\bar t$, 
where $m_{S^0} > 2m_t$ ($S^0 = H^0,\, A^0$) was assumed.
On one hand this was because $gg \to S^0 \to t\bar t$
may be swamped by interference effect with a much larger QCD-produced $t\bar t$.
On the other hand, there was an implicit expectation that
the case for $A^0$ below $t\bar t$ threshold is likely ruled out already.
However, as demonstrated above, the lighter $A^0$ case is
not yet ruled out, even though current data can already cut into
the parameter space.

This scenario, where one can turn off processes such as
$gg \to A^0$ by suppressing $\rho_{tt}$,
or $A^0 \to h^0Z$ by small $\cos\gamma$ (alignment),
is rather definite in the $cg \to tA^0 \to tt\bar c,\ t\bar tc$ final state,
with $A^0 \to t\bar c + \bar t c$ close to 100\%
while evading $A^0 \to h^0Z$ searches.
We also put forward the important point that
the small $\rho_{tt}$ case may call on $\rho_{tc} = {\cal O}(1)$
for electroweak baryogenesis~\cite{Fuyuto:2017ewj}, 
providing thus a strong motivation beyond just searching for exotic scalar bosons.
We urge the ATLAS and CMS experiments to pursue
both same-sign top and triple top, and in particular the
same-sign top coming from a relatively light $A^0$ in $tA^0$ associated production.

Of course, $cg \to tS^0 \to t t \bar t$~\cite{Kohda:2017fkn} triple top production
provides a rather attractive signature, as the SM processes are rather suppressed.
It can also differentiate, in the longer term, between the lighter and heavier scalar 
scenarios if a precursor hint for same-sign top appears.


Insight to FCNH $tch^0$ coupling started with $t \to ch^0$ \cite{Hou:1991un, Chen:2013qta},
which has not been discovered so far.
This is consistent with approximate alignment,
that the $CP$ even $h^0$-$H^0$ scalar mixing angle $\cos\gamma$ is small.
It is interesting to note that approximate alignment may be closely
associated~\cite{Hou:2017hiw} with ${\cal O}(1)$ Higgs quartic couplings,
which is needed for first order EWPT.
But how do we access the exotic Higgs sector if $\cos\gamma$ is very small?
The same-sign top signatures studied here for $A^0$ below $t\bar t$ threshold, 
as well as the heavier scalar case studied in Ref.~\cite{Kohda:2017fkn}, 
provide entry points for LHC access.

But $\cos\gamma$ may not be that small. Current studies~\cite{cos_b-a}
assume $Z_2$ symmetry, while a dedicated study in the general 2HDM without $Z_2$,
with many more parameters, is still lacking.
If $\cos\gamma$ is not vanishingly small, we see from Fig.~1
that a finite $A^0 \to h^0Z$ branching ratio brings about
$cg \to tA^0 \to tZh^0$, which is a striking final state.
This process would be studied in a separate work.
However, given that
 ${\cal B}(A^0 \to h^0Z) {\cal B}(Z \to \ell^+\ell^-)
 \ll {\cal B}(A^0 \to t\bar c) {\cal B}(t \to b\ell^+\nu_\ell)$,
the expected cross section would be much less than 
the fb level in Table I.
We remark in passing that a finite $\cos\gamma$ may
allow us to probe~\cite{Hou:2018uvr} the extra Yukawa coupling 
$\rho_{tt}$  via interference in the recently observed~\cite{Sirunyan:2018hoz, Aaboud:2018urx} 
$pp \to t\bar t h^0$ production process.

In studying $cg \to tA^0$, we have assumed 
$A^0$ to be considerably lighter than $H^0$ and $H^+$, 
forbidding $A^0 \to H^0Z$ and $H^\pm W^\mp$ kinematically.
To allow $A^0$ to be lighter but to reduce tension with 
electroweak precision $S$ and $T$ constraints, 
$H^0$--$H^+$ would likely be close to degenerate, 
which may be justified by the so-called twisted custodial symmetry~\cite{Gerard:2007kn}.
Alternatively, one can consider $cg \to tH^0$ assuming a case where $H^0$ is lighter
with the custodial symmetry, or near degeneracy of $A^0$--$H^+$.
In particular, similar to our discussion for $A^0$,
with $H^0\to A^0Z$, $H^\pm W^\mp$ kinematically forbidden,
one could have $H^0 \to t\bar c,\, \bar tc$ dominance in the alignment limit.
With approximate alignment, $H^0 \to h^0h^0$ may be less suppressed than
$A^0 \to h^0Z$, leading to the ``top-tagged di-Higgs'' signature,
i.e. $cg \to tH^0 \to th^0h^0$, which would be studied elsewhere.

We have studied in Ref.~\cite{Kohda:2017fkn}, above $t\bar t$ threshold,
the case where $A^0$ and $H^0$ are close in mass,
and found that $cg \to tA^0 \to tt\bar{c}$ and $cg\to tH^0 \to tt\bar{c}$ amplitudes
cancel each other up to difference in mass and width for $c_\gamma =\rho_{ct}=0$.
But if $m_{A^0}$-$m_{H^0}$ splitting is larger than the widths involved, 
the interference effect is diminished.
As an illustration for our present case,  
for $m_{A^0}=300$ GeV, $m_{H^0} \sim 330$ GeV and $\rho_{tc}= 1$, 
which imply widths $\Gamma_{A^0} \simeq 8$ GeV, $\Gamma_{H^0} \simeq 10$ GeV
[$\mathcal{B}(A^0, H^0 \to t\bar c, \bar t c) \simeq 100$\% assumed],
the interference effect is smaller than 10\% 
of the $cg \to tt\bar c$ cross section, which is in fact larger than $A^0$ alone
because of the incoherent $H^0$ contribution.
This implies both tighter exclusion and better discovery reach
for $\rho_{tc}$ than Fig.~\ref{exclusion}, up to a factor of $\sim \sqrt{2}$.
But the effect of $H^0$ becomes negligible for large enough $m_{A^0}$-$m_{H^0}$ splitting.
For example, for $m_{A^0} = 300$ GeV and $m_{H^0} = 550$ GeV, 
the $H^0$ contribution is $\sim 20\%$ of the $A^0$ one,
implying $\sim 10\%$ change to $|\rho_{tc}|$ values in Fig.~\ref{exclusion}.
Such $m_{H^0}$ values imply a mildly large Higgs quartic coupling,
$\eta_5 \simeq (m_{H^0}^2 -m_{A^0}^2)/v^2 \simeq 3.5$
in the notation of Ref.~\cite{Hou:2017hiw}.
Even larger mass splittings require larger $\eta_5$, 
which would eventually run into trouble with perturbativity.

In conclusion,
it is mildly surprising that a relatively light $A^0$ below $t\bar t$ threshold
that possess FCNH coupling $\rho_{tc}$, is not yet ruled out.
With $gg \to A^0$ suppressed by small $\rho_{tt}$, the extra diagonal top Yukawa coupling, 
and with $A^0 \to h^0Z$ further suppressed by approximate alignment,
one could have $cg \to tA^0$ followed by $A^0 \to t\bar c + \bar tc$ at 100\%,
leading to same-sign top signature that can be studied at the LHC.
The byproduct of an existing CMS study already cuts into the
parameter space of $\rho_{tc}$-lead electroweak baryogenesis,
while the whole program can be (almost) fully probed with
3000 fb$^{-1}$ data.

\vskip0.2cm
\noindent{\bf Acknowledgments} \
We thank K.-F. Chen and Y.~Chao for fruitful discussions.
This research is supported by grants MOST 104-2112-M-002-017-MY3,
107-2811-M-002-039, and 106-2811-M-002-187.



\begin{thebibliography}{99}
\bibitem{h125}
  We assume $h^0$ corresponds to the observed 125 GeV boson,
  with $m_{h^0} < m_{H^0}$.
\end{thebibliography}
\end{document}